\documentclass[preprint,onecolumn,nofootinbib]{revtex4}
\usepackage[colorlinks=true,linkcolor=blue,urlcolor=blue,filecolor=black,citecolor=red,pdfstartview=FitV,pdftitle={},pdfsubject={},pdfkeywords={},pdfpagemode=None,bookmarksopen=true]{hyperref}
\usepackage{graphicx}
\usepackage{amsmath}
\usepackage{amsfonts}
\usepackage{amssymb,ulem}
\usepackage{color,xcolor}%
\usepackage{CJK}
\usepackage{subfigure}
\usepackage{amsthm,amsmath,amssymb}
\usepackage{mathrsfs}
\usepackage{multirow}

\usepackage{dcolumn}
\usepackage{float}
\begin{document}
\title{Assessing EMRI Detectability of the Rotating Quantum Oppenheimer-Snyder Black Hole }
\author{Dan Zhang $^{1}$}
\thanks{danzhanglnk@163.com}
\author{Shulan Li$^{2}$}
\thanks{shulanli.yzu@gmail.com}
\author{Guoyang Fu$^{1}$}
\thanks{fuguoyang@yzu.edu.cn}
\author{Jian-Pin Wu$^{1}$}
\thanks{jianpinwu@yzu.edu.cn, corresponding author}
\affiliation{
$^1$\mbox{Center for Gravitation and Cosmology, College of Physical Science and Technology,} \mbox{Yangzhou University, Yangzhou 225009, China}\\
$^{2}$\mbox{Department of Physics, Shanghai University, Shanghai 200444, China}
}

\begin{abstract}
This letter presents an assessment of quantum gravity effects on extreme-mass-ratio inspirals (EMRIs) for the rotating quantum Oppenheimer-Snyder (qOS) black hole. Employing the adiabatic evolution, we compute the gravitational wave (GW) dephasing, which quantifies the cumulative phase shift induced by the quantum correction $\alpha$. We further generate the augmented analytic kludge (AAK) waveform and investigate the faithfulness between the waveforms with and without the quantum parameter $\alpha$ for different values of $a$. Our results reveal that the quantum gravity effect induces detectable imprints in LISA, while the presence of rotation suppresses these signatures. This suggests that rotational degrees of freedom must be carefully accounted for when probing quantum gravity with EMRI observations.

\end{abstract}

\maketitle

\section{Introduction}  \label{sec-intro}

Classical general relativity (GR), despite its remarkable success in describing gravitational phenomena across a wide range of scales, predicts the inevitable emergence of spacetime singularities, regions where the theory itself breaks down \cite{Penrose:1964wq, Hawking:1970zqf}. Resolving these singularities is widely regarded as a key task for a consistent theory of quantum gravity. Among various proposals, loop quantum gravity (LQG) has emerged as a compelling candidate, distinguished by its background-independent and non-perturbative formulation \cite{Rovelli:1997yv, Thiemann:2002nj, Ashtekar:2004eh, Han:2005km}. A notable achievement of LQG is its ability to resolve the Big Bang singularity in cosmology \cite{Bojowald:2001xe, Ashtekar:2006rx, Ashtekar:2013hs}. This framework has since been successfully extended to static spherically symmetric black hole (BH), where the classical singularity is replaced by a regular transition surface or a quantum bounce(see, e.g., Ref.~\cite{Peltola:2008pa, Peltola:2009jm, Modesto:2008im, Ashtekar:2018lag, Ashtekar:2018cay, Gambini:2020nsf, Bodendorfer:2019cyv, Bodendorfer:2019nvy, Kelly:2020uwj, Parvizi:2021ekr, Lewandowski:2022zce, Giesel:2022rxi, Alonso-Bardaji:2021yls, Alonso-Bardaji:2022ear, Zhang:2024khj} and references therein)
More recently, a framework that effectively incorporates LQG effects was recently employed to investigate the gravitational collapse of a dust ball \cite{Lewandowski:2022zce}. That study indicates that as the energy density approaches the Planck scale, the collapse halts and a bounce occurs inside the dust ball. It also presents a new quantum-corrected Schwarzschild metric, henceforth referred to as quantum Oppenheimer--Snyder (qOS) BH model.

To more realistically model astrophysical BHs, it is necessary to extend the static solution to a rotating counterpart, a crucial step for connecting LQG with observations. 
However, constructing exact rotating solutions by directly loop quantizing axisymmetric spacetimes currently remains quite challenging. A practical alternative is to employ the Newman-Janis algorithm (NJA) \cite{Newman:1965tw} or its refined version \cite{Azreg-Ainou:2014pra, Azreg-Ainou:2014aqa}, which generates an axisymmetric metric from a static spherically symmetric seed.
Despite the lack of a rigorous justification for applying the NJA beyond GR, the resulting rotating metrics serve as effective phenomenological models that capture key features expected from quantum-corrected geometries. This approach has been widely adopted to generate rotating LQG BH solutions from various spherically symmetric seeds \cite{Caravelli:2010ff, Liu:2020ola, Brahma:2020eos, Chen:2022nix, Kumar:2022vfg, Huang:2022iwl, Fazzini:2024nzq, Ali:2024ssf, Ban:2024qsa}.
Notably, the rotating version of the qOS BH has been constructed via the revised NJA and studied in several recent works, with topics covering BH shadows and strong gravitational lensing effects together with the associated constraints from Event Horizon Telescope (EHT) observations \cite{Ali:2024ssf, Vachher:2024ait, Raza:2025ohk}, the Penrose process and energy extraction \cite{Fatima:2025fdc}, as well as quasinormal modes \cite{Chen:2025wfi}. These studies collectively demonstrate that the rotating quantum-corrected BH model provides a viable framework for probing quantum gravity signatures using current and upcoming astrophysical observations.

The discovery of gravitational waves (GWs) has opened a new observational window that may ultimately provide access to quantum gravity effects \cite{LIGOScientific:2016emj,LIGOScientific:2017vwq,LIGOScientific:2018mvr,LIGOScientific:2020ibl,LIGOScientific:2021usb,KAGRA:2021vkt}. Although no definitive evidence of quantum gravity has yet been found in GW data, the next generation of space-based detectors, such as LISA \cite{LISA:2017pwj,LISA:2024hlh}, TianQin \cite{Luo:2015ght,TianQin:2020hid,Gong:2021gvw}, and Taiji \cite{Hu:2017mde,Gong:2021gvw}, will test GR with higher precision and are expected to capture characteristic signals of quantum gravity. Extreme mass-ratio inspirals (EMRIs), where a stellar-mass compact object inspirals into a supermassive BH, are primary scientific targets for these future detectors \cite{Babak:2017tow,Gair:2004iv,Mapelli:2012pw}. EMRIs are highly sensitive to the spacetime geometry \cite{Amaro-Seoane:2007osp,Barack:2006pq}, such that even tiny effects accumulate over time and become observationally detectable. This makes them a powerful tool for probing quantum gravity effects \cite{Fu:2024cfk,Zi:2024jla,Liu:2024qci,Yang:2025esa,Ahmed:2025shr,Gong:2025mne,Chen:2026kbn,Zhang:2026lrd}.

Recently, the authors of Ref.~\cite{Yang:2025esa} investigated the potential of probing quantum gravity effects with EMRIs in a static qOS background. However, the absence of rotational effects may bias the assessment. In this letter, we extend their analysis to the rotating case and investigate the imprints of quantum gravity effects on EMRIs in rotating qOS background. The remainder of this letter is structured as follows. In Sec. \ref{sec-1}, we present the rotating qOS BH model and analyze its horizon structure. In Sec. \ref{sec-2}, we calculate the adiabatic orbital evolution and investigate the effects of the quantum parameter $\alpha$ and the rotational parameter $a$ on dephasing. In Sec. \ref{sec-3}, we generate the EMRI waveform using the \texttt{FEW} package and evaluate the distinguishability of quantum gravity effects via faithfulness. Finally, we summarize our results in Sec. \ref{sec-4}.

\section{Rotating Quantum Oppenheimer-Snyder Model}\label{sec-1}
In this section, we briefly review the rotating version of the qOS BH model. The static seed metric is given by \cite{Lewandowski:2022zce}
\begin{eqnarray} \label{metric_sph}
\mathrm{d}s^2=-f(r)\mathrm{d}t^2+f(r)^{-1}\mathrm{d}r^2 +r^2(\mathrm{d}\theta^2+\sin^2\theta \mathrm{d}\phi^2)\,,
\end{eqnarray}
where
\begin{eqnarray} \label{metric_fun}
f(r)=1-\frac{2M}{r}+\frac{\alpha M^2}{r^4}.
\end{eqnarray}
Hence, the static qOS model offers an effective quantum gravity description by introducing the LQG correction term $\alpha M^2/r^4$, where $\alpha$ quantifies the deviation from the Schwarzschild case. Notably, the quantum-corrected parameter $\alpha$ must be satisfy $\alpha/M^2\leq 27/16$ to ensure the existence of an event horizon.

By applying the revised NJA \cite{Azreg-Ainou:2014pra, Azreg-Ainou:2014aqa}, the static qOS metric \eqref{metric_sph} with \eqref{metric_fun} is extended to the rotating counterpart, yielding the following line element in Boyer-Lindquist coordinates \cite{Ali:2024ssf, Vachher:2024ait, Raza:2025ohk, Fatima:2025fdc, Chen:2025wfi}
\begin{eqnarray} \label{metric_rot}
\mathrm{d} s^{2}&= & -\left(\frac{\Delta-a^{2} \sin ^{2} \theta}{\Sigma}\right) \mathrm{d} t^{2}-2 a \sin ^{2} \theta\left(1-\frac{\Delta-a^{2} \sin ^{2} \theta}{\Sigma}\right) \mathrm{d} t \mathrm{~d} \phi \nonumber \\ 
&& +\sin ^{2} \theta\left[\Sigma+a^{2} \sin ^{2} \theta\left(2-\frac{\Delta-a^{2} \sin ^{2} \theta}{\Sigma}\right)\right] \mathrm{d} \phi^{2}+\frac{\Sigma}{\Delta} \mathrm{d} r^{2}+\Sigma \mathrm{d} \theta^{2},
\end{eqnarray}
where $\Delta=r^2+a^2-2M r + \alpha M^2/r^2$ and $\Sigma=r^2+a^2\cos^2\theta$. The metric \eqref{metric_rot} reduces to the qOS BH and Kerr BH in the limits $a\to0$ and $\alpha\to0$, respectively.  

\begin{figure}[H]
	\centering
	\includegraphics[width=0.55\textwidth]{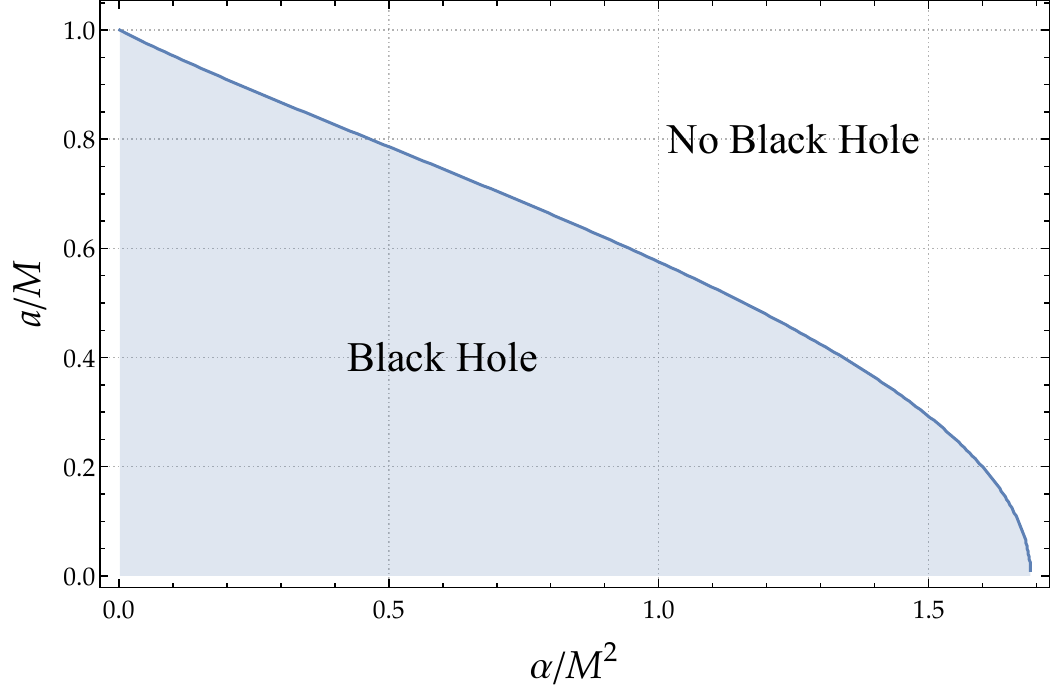}\hspace{0.2mm}	
	\caption{Phase diagram of the rotating qOs model in the parameter space of $a/M$ and $\alpha/M^2$. The blue line corresponds to the extremal BH.}
	\label{phase_diagram}
\end{figure}

The horizon of a rotating qOS BH is determined by the dimensionless parameters $a/M$ and $\alpha/M^2$, with its location given by the condition $\Delta=0$, as shown in Fig. \ref{phase_diagram}.  The blue shaded region corresponds to the existence of two horizons, while the blue line marks their degenerate case, representing an extremal BH. For parameters above this line, the horizon vanishes, corresponding to a no BH scenario. In the following, we adopt dimensionless parameters $\hat{a}=a/M$ and $\hat{\alpha}=\alpha/M^2$. For convenience, the hats on these dimensionless variables are omitted hereafter. 

\section{Adiabatic Orbital Evolution}\label{sec-2}
EMRI describes the bound orbital evolution of a stellar mass object driven by gravitational radiation reaction. Since the timescale of radiation reaction is much longer than the orbital period, this allows us to separate the problem into two steps: first, computing the geodesic of the stellar-mass object; and second, calculating the radiation reaction fluxes. In this section, we investigate the trajectory of the stellar-mass object and analyze the impact of quantum effects on the dephasing.

We treat the stellar-mass object as a point particle and focus on equatorial orbits ($\theta=\pi/2$). There are two Killing vectors $\xi^\mu_{(t)}=(1,0,0,0)$ and $\xi^\mu_{(\phi)}=(0,0,0,1)$, corresponding to the energy $E$ and angular momentum $L_z$, which can be written as
\begin{eqnarray}
\label{4_velocity}
-1&=&g_{\mu\nu}u^\mu u^\nu \\ 
\label{En}
E&=&-u_\mu \xi^\mu_{(t)}=-g_{tt}u^t-g_{t\phi}u^\phi , \\
\label{Lz}
L_z&=&u_\mu \xi^\mu_{(\phi)}=g_{t\phi}u^t+g_{\phi\phi}u^\phi,
\end{eqnarray}
where $u^\mu$ denotes the $4$-velocity. For eccentric orbits, the trajectory is characterized by the semi-latus rectum $p$ and eccentricity $e$ as 
\begin{eqnarray}
e=\frac{r_a-r_p}{r_a+r_p}, \ p=\frac{2r_ar_p}{r_a+r_p},
\end{eqnarray}
where the $r_a$ and $r_p$ represent the orbit's apastron and periastron, respectively. For given orbital parameters $\{p,e\}$, one can numerically solve for $\{E,L_z\}$ using the condition $dr/d\tau=0$ at $r_a$ and $r_p$. To interpolate more efficiently near the strong field region, we adopt the uniform grids $\{u,e\}$, where $u=\log(p-p_s+3.9)$ and $p_s$ is the separatrix of the Kerr case. This grid allows us to concentrate more grid points near the separatrix \cite{Chua:2020stf,Katz:2021yft}.

For the bound orbital evolution with the radiation reaction, the GW fluxes can be obtained from the quadrupole formula \cite{Thorne:1980ru,Ryan:1995zm,Peters:1963ux}
\begin{eqnarray}
\label{E_flux}
\left<\dot{E}\right>&=&-\frac{1}{5}\left<I_{TT}^{jk(3)}I_{TT}^{jk(3)}+\frac{16}{9}J_{TT}^{jk(3)}J_{TT}^{jk(3)}\right>, \\
\label{Lz_flux}
\left<\dot{L}_z\right>&=&-\frac{2}{5}\epsilon^{3kl}\left<I_{TT}^{ka(3)}I_{TT}^{al(3)}+\frac{16}{9}J_{TT}^{ka(3)}J_{TT}^{al(3)}\right>. 
\end{eqnarray}
The mass and current quadrupole moments are defined as
\begin{eqnarray}
I^{jk}&=&\mu x^j x^k, \\
J^{jk}&=&\epsilon_{jml} v^m I^{lk}.
\end{eqnarray}
In order to obtain the fluxes on a uniform grid, we numerically integrate \eqref{E_flux} and \eqref{Lz_flux} over the quasi-Keplerian true anomaly $\chi$ from $0$ to $2\pi$, and then perform a bicubic spline interpolation on the $\{u,e\}$ grid. 

Once the fluxes are known, the evolution of the orbital parameters under the adiabatic approximation is governed by \cite{Cutler:1994pb, Glampedakis:2002ya}
\begin{eqnarray}
\frac{dp}{dt}&=&\left[-\frac{\partial L_z}{\partial e}\dot{E}+\frac{\partial E}{\partial e}\dot{L}_z\right]\bigg/\left[\frac{\partial L_z}{\partial e}\frac{\partial E}{\partial p}-\frac{\partial E}{\partial e}\frac{\partial L_z}{\partial p}\right],\\
\frac{de}{dt}&=&\left[\frac{\partial L_z}{\partial p}\dot{E}-\frac{\partial E}{\partial p}\dot{L}_z\right]\bigg/\left[\frac{\partial L_z}{\partial e}\frac{\partial E}{\partial p}-\frac{\partial E}{\partial e}\frac{\partial L_z}{\partial p}\right], 
\end{eqnarray}
where $\left<\dot{E}\right>=-\mu\dot{E}$ and $\left<\dot{L}_z\right>=-\mu \dot{L}_z$. The phase $\{\Phi_{\phi},\Phi_r\}$ is an important orbital parameter for quantifying possible GW detections by future space-based GW detector such as LISA, and it can be obtained by integrating the orbital frequency over time 
\begin{eqnarray}\label{GWphase}
\Phi_i=\int\Omega_i(p(t),e(t))dt , \  i=\phi, \ r. 
\end{eqnarray}
To quantify the effect of the quantum parameter $\alpha_0$, we define dephasing by the phase after subtracting the contribution from GR
\begin{eqnarray}\label{GWphase}
\Delta\Phi_i=\Phi_i^{\alpha}-\Phi_i^{\alpha=0}, \ i=\phi, \ r
\end{eqnarray}

\begin{figure}[H]
	\centering
	\includegraphics[width=0.55\textwidth]{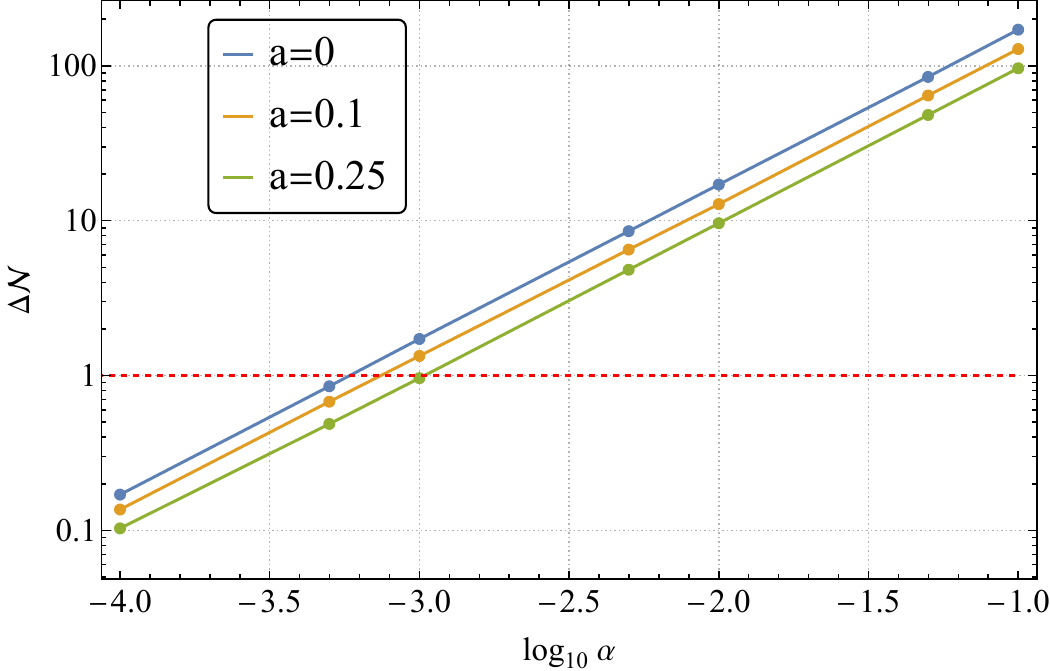}\hspace{0.2mm}	
	\caption{Dependence of the dephasing on the quantum parameter $\alpha$ for different values of $a$. The red dashed line indicates the detection threshold of $\Delta\Phi\sim0.1$ rad.}
	\label{dephasing}
\end{figure}

Since the contribution from radial dephasing is much smaller than that from the azimuthal component, i.e., $\Delta\Phi_r\ll \Delta\Phi_\phi$, the dephasing of the GW can be approximated by $\Delta\Phi\sim\Delta\Phi_\phi$. For fixed initial semi-latus rectum $p_0 = 9.8$, eccentricity $e_0 = 0.1$, central BH mass $M = 10^6 M_\odot$, and stellar-mass object mass $\mu = 10 M_\odot$, we plot the dephasing as a function of $\alpha$ for different values of $a$ in Fig. \ref{dephasing}.  
Following \cite{Gupta:2021cno,Bonga:2019ycj}, we set $\Delta\Phi\sim1$ rad as the criterion for detectability, indicated by the red dashed line in Fig. \ref{dephasing}. It is evident that, after one-year accumulation, the dephasing becomes more pronounced as the quantum parameter increases. However, when rotation is taken into account, the dephasing decreases for a fixed parameter $\alpha$. This indicates that the presence of rotation suppresses the dephasing of GWs arising from quantum effects.  

\section{Waveform and Faithfulness}\label{sec-3}

In this section, we investigate the EMRI waveform and further assess the distinguishability of quantum gravity effects. The kludge waveform, one of the most important EMRI waveforms, is widely employed in LISA data analysis because of the advantage of rapid generation \cite{Chua:2017ujo,PhysRevD,Chua:2015mua}. Based on the original augmented analytic kludge (AAK) framework, a new version of the augmented analytic kludge (AAK) waveform has been developed by the \texttt{FastEMRIWaveforms (FEW)} \cite{Katz:2021yft}, which directly computes fundamental frequencies along the orbital trajectory, thereby eliminating the need for frequency mapping (see [83,84] and references therein for technical details). 

\begin{figure}[H]
	\centering
	\includegraphics[width=0.45\textwidth]{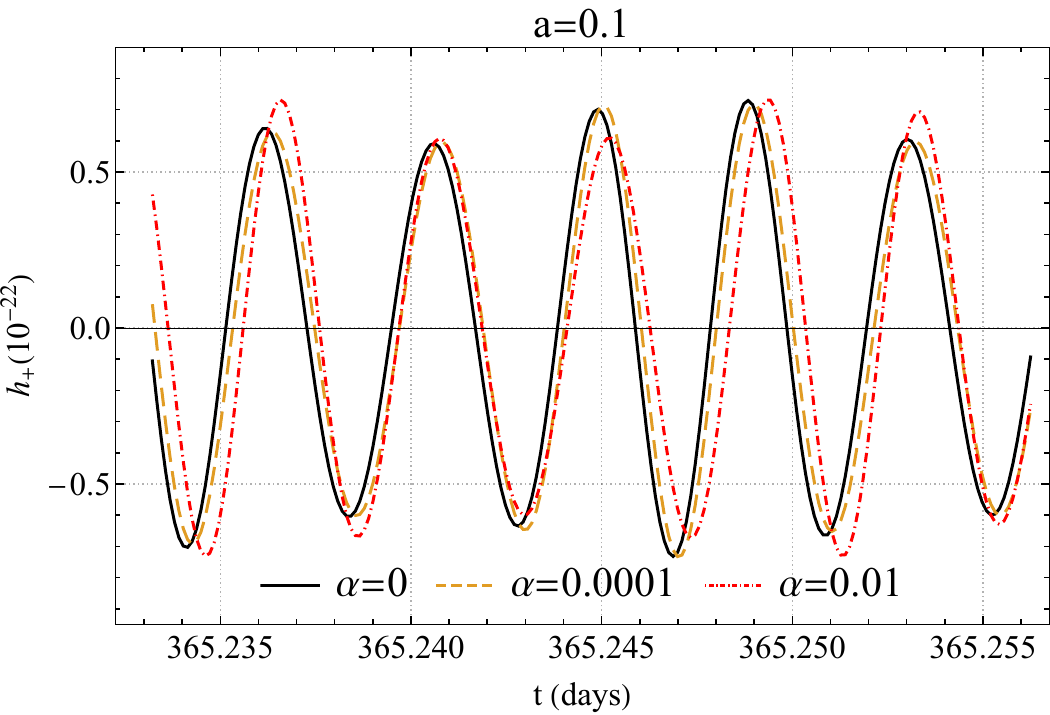}\hspace{0.2mm}	
    \includegraphics[width=0.45\textwidth]{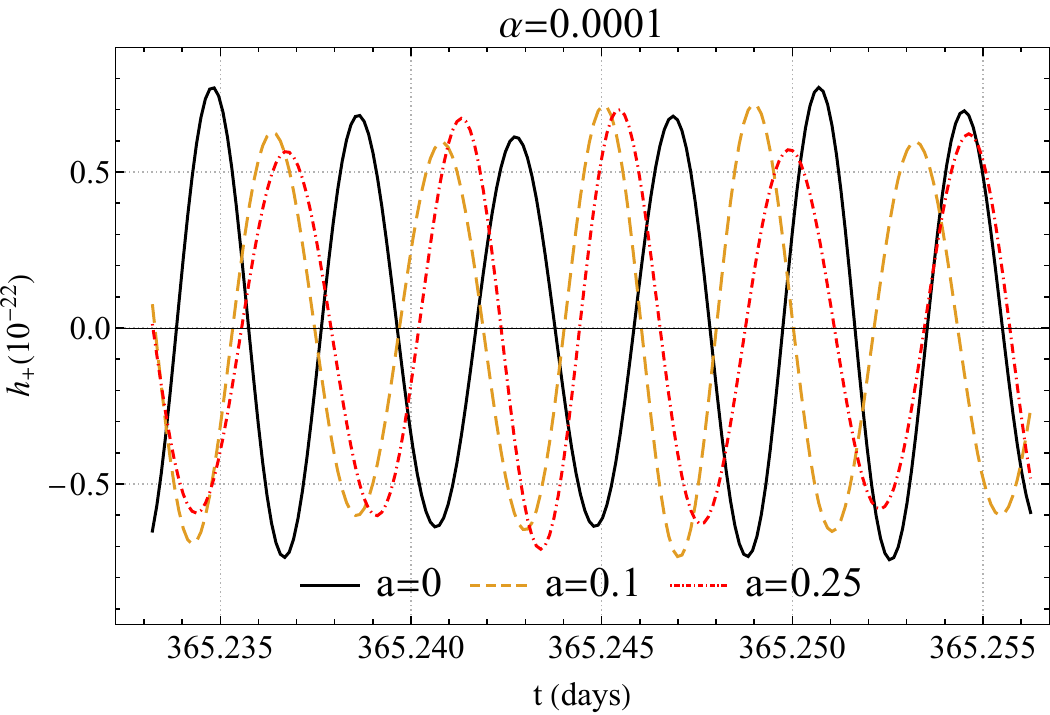}\hspace{0.2mm}	
	\caption{The AAK waveform for the rotation qOS background with different $a$ and $\alpha$.}
	\label{AAK_waveform}
\end{figure}

Assuming the same initial conditions, we generate the AAK waveform for the rotating qOS model using the \texttt{FEW} package. Fig. \ref{AAK_waveform} exhibits the AAK waveforms after one-year of evolution for different values of the parameters $a$ and $\alpha$. For a fixed $a=0.1$ (left panel of Fig. \ref{AAK_waveform}), we observe differences in the GW waveform between GR and the rotating qOS model due to quantum gravity effects. Moreover, the right panel of Fig. \ref{AAK_waveform} shows that as $a$ increases, rotational effects drive the EMRI waveform to deviate more significantly from the GR. This implies that rotational effects play a crucial role and cannot be neglected when modeling quantum-corrected EMRI waveforms.

For a more accurate assessment of the differences in EMRI waveforms, we compute the faithfulness $\mathcal{F}$ between GW signals with and without quantum gravity correction. It is defined as
\begin{eqnarray}\label{faithfulness}
\mathcal{F}\left[h_{1}, h_{2}\right]=\max _{\left\{t_{c}, \phi_{c}\right\}} \frac{\left\langle h_{1} \mid h_{2}\right\rangle}{\sqrt{\left\langle h_{1} \mid h_{1}\right\rangle\left\langle h_{2} \mid h_{2}\right\rangle}}\,,
\end{eqnarray}
where the noise-weighted inner product is given by
\begin{equation}
\langle h_1 \mid h_2 \rangle = 4 \Re \int_{f_{min}}^{f_{max}} \frac{\tilde{h}_1(f) \tilde{h}_2^*(f)}{S_n(f)} df,
\end{equation}
with $\tilde{h}(f)$ representing the Fourier transform of $h(t)$, and $S_n(f)$ the power spectral density of LISA, including the confusion noise from galactic white dwarf binaries \cite{Robson:2018ifk}. Following Refs.\cite{Chatziioannou:2017tdw}, LISA can distinguish two signals when the faithfulness satisfies $\mathcal{F}\lesssim 1-d/(2\text{SNR})^2$. Without loss of generality, we set $\mathcal{F}\simeq0.965$ as the criterion to distinguish between the two signals.   

Fig.~\ref{faithfulness} presents the faithfulness as a function of the quantum parameter $\alpha$ for different values of $a$. As expected, as $\alpha$ increases, quantum gravity effects cause the EMRI waveform to deviate more significantly from GR, and consequently the faithfulness decreases. For a fixed quantum parameter $\alpha$, the faithfulness $\mathcal{F}$ increases with $a$, indicating that rotational effects tend to suppress the distinguishability of quantum gravity imprints in the EMRI waveform. This observation implies that rotational degrees of freedom must be carefully accounted for when assessing the detectability of quantum gravity effects with LISA.

\begin{figure}[H]
	\centering
	\includegraphics[width=0.55\textwidth]{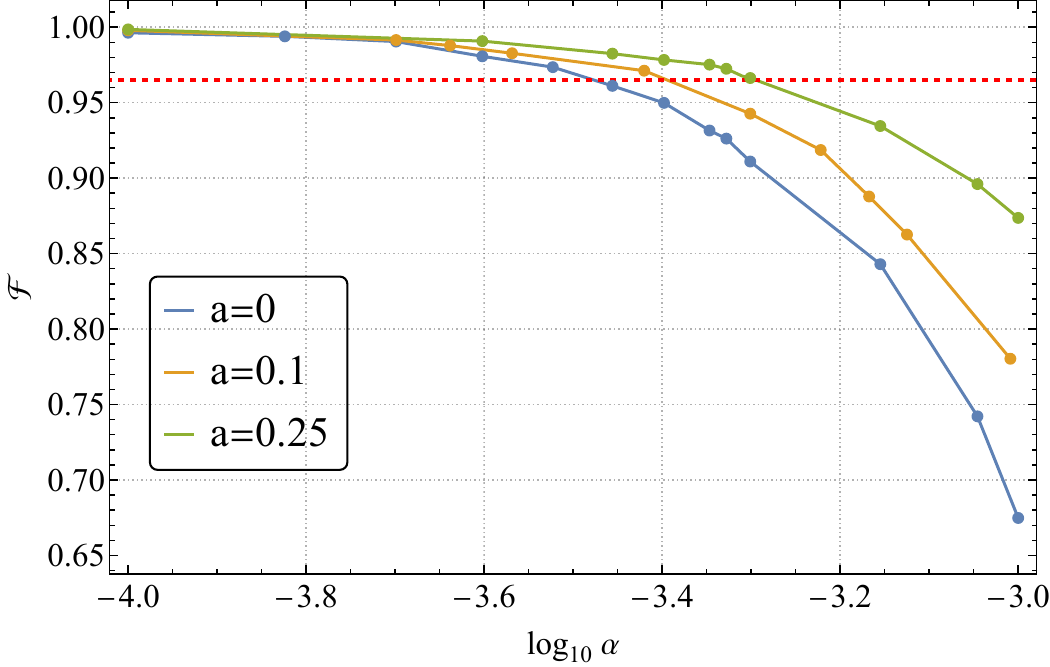}\hspace{0.2mm}	
	\caption{Faithfulness between the with and without quantum gravity correction with varying $a$. The red dashed line denotes the distinguishability criterion $\mathcal{F}\simeq0.965$.}
	\label{faithfulness}
\end{figure}

\section{Conclusions}\label{sec-4}

In this letter, we adopt the rotating qOS BH model and systematically investigate how the quantum parameter $\alpha$ and the rotational parameter $a$ affect the observable signatures of EMRIs. When the weak-field approximation is abandoned, the energy flux and orbital evolution are rendered not analytically expressible due to the inability to expand in large $p$. To address this problem, we perform numerical integration with bicubic splines over the $\{u,e\}$ grid. We then compute the adiabatic orbital evolution and analyze the dephasing induced by the quantum parameter $\alpha$ for different values of $a$. Our results reveal that, as time accumulates, the dephasing induced by quantum corrections becomes detectable, while the presence of rotation suppresses this effect.

Furthermore, we generate the AAK waveform using the \texttt{FEW} package and find that waveforms with different parameters exhibit significant differences after one year of evolution. To further quantify this difference, we calculate the faithfulness between waveforms with and without quantum corrections. The faithfulness decreases with increasing $\alpha$ but increases with $a$ for fixed $\alpha$, indicating that rotational effects tend to mask the imprints of quantum gravity. Our analysis highlights the importance of incorporating rotation in quantum-corrected BH models for reliable GW predictions.

\acknowledgments

This work is supported by National Key R$\&$D Program of China (No. 2023YFC2206703), the Natural Science Foundation of China under Grant Nos. 12275079, 12505078, 12505085, 12375055, the China Postdoctoral Science Foundation (Grant No. 2025T180931), and the Jiangsu Funding Program for Excellent Postdoctoral Talent (Grant No. 2025ZB705).


\bibliographystyle{style1}
\bibliography{Ref}
\end{document}